\documentclass[12pt]{article}
\usepackage{epsf,latexsym}
\newcommand{\gsim}{\;\rlap{\lower 3.5 pt \hbox{$\mathchar \sim$}} \raise 1pt
 \hbox {$>$}\;}
\newcommand{\lsim}{\;\rlap{\lower 3.5 pt \hbox{$\mathchar \sim$}} \raise 1pt
 \hbox {$<$}\;}
\newcommand{\re}{\mathop{\mathrm{Re}}\nolimits}
\newcommand{\tr}{\mathop{\mathrm{tr}}\nolimits}
\newcommand{\li}{\mathop{\mathrm{Li}_2}\nolimits}

\catcode`@=11
\newcount\@tempcntc
\def\@citex[#1]#2{\if@filesw\immediate\write\@auxout{\string\citation{#2}}\fi
  \@tempcnta\z@\@tempcntb\m@ne\def\@citea{}\@cite{\@for\@citeb:=#2\do
    {\@ifundefined
       {b@\@citeb}{\@citeo\@tempcntb\m@ne\@citea\def\@citea{,}{\bf
?}\@warning
       {Citation `\@citeb' on page \thepage \space undefined}}%
    {\setbox\z@\hbox{\global\@tempcntc0\csname b@\@citeb\endcsname\relax}%
     \ifnum\@tempcntc=\z@ \@citeo\@tempcntb\m@ne
       \@citea\def\@citea{,}\hbox{\csname b@\@citeb\endcsname}%
     \else
      \advance\@tempcntb\@ne
      \ifnum\@tempcntb=\@tempcntc
      \else\advance\@tempcntb\m@ne\@citeo
      \@tempcnta\@tempcntc\@tempcntb\@tempcntc\fi\fi}}\@citeo}{#1}}
\def\@citeo{\ifnum\@tempcnta>\@tempcntb\else\@citea\def\@citea{,}%
  \ifnum\@tempcnta=\@tempcntb\the\@tempcnta\else
   {\advance\@tempcnta\@ne\ifnum\@tempcnta=\@tempcntb \else
\def\@citea{--}\fi
    \advance\@tempcnta\m@ne\the\@tempcnta\@citea\the\@tempcntb}\fi\fi}
\catcode`@=12

\begin{document}
\title{\vskip-3cm{\baselineskip14pt
\centerline{\normalsize CERN-TH/2002-095\hfill ISSN 0418-9833}
\centerline{\normalsize DESY 02-057\hfill}
\centerline{\normalsize hep-ph/0205312\hfill}
\centerline{\normalsize May 2002\hfill}}
\vskip1.5cm
QCD Corrections to $t\overline{b}H^-$ Associated Production in $e^+e^-$
Annihilation}
\author{{\sc Bernd A. Kniehl}\thanks{Permanent address: II. Institut f\"ur
Theoretische Physik, Universit\"at Hamburg, Luruper Chaussee 149, 22761
Hamburg, Germany.}\\
{\normalsize CERN, Theoretical Physics Division, 1211 Geneva 23, Switzerland}\\
\\
{\sc Fantina Madricardo, Matthias Steinhauser}\\
{\normalsize II. Institut f\"ur Theoretische Physik, Universit\"at
Hamburg,}\\
{\normalsize Luruper Chaussee 149, 22761 Hamburg, Germany}}

\date{}

\maketitle

\thispagestyle{empty}

\begin{abstract}
We calculate the QCD corrections to the cross section of
$e^+e^-\to t\overline{b}H^-$ and its charge-conjugate counterpart within the
minimal supersymmetric extension of the Standard Model.
This process is particularly important if $m_t<m_H+m_b$ and $\sqrt s<2m_H$, so
that $t\to bH^+$ and $e^+e^-\to H^+H^-$ are not allowed kinematically.
Large logarithmic corrections that arise in the on-mass-shell scheme of quark
mass renormalization, especially from the $t\overline{b}H^-$ Yukawa coupling
for large values of $\tan\beta$, are resummed by adopting the modified
minimal-subtraction scheme, so that the convergence behavior of the
perturbative expansion is improved.
The inclusion of the QCD corrections leads to a significant reduction of the
theoretical uncertainties due to scheme and scale dependences.

\medskip

\end{abstract}

\newpage

\section{Introduction}

One of the prime objectives of a future $e^+e^-$ linear collider (LC) will be
the detailed study of spin-zero particles which remain in the physical
spectrum after the elementary-particle masses have been generated through the
Higgs mechanism of electroweak symmetry breaking.
Should the world be supersymmetric, then the Higgs sector is more complicated
than in the Standard Model (SM), which predicts just one neutral CP-even Higgs
boson $H$.
The Higgs sector of the minimal supersymmetric extension of the SM (MSSM)
consists of a two-Higgs-doublet model (2HDM) and accommodates five physical
Higgs bosons:
the neutral CP-even $h^0$ and $H^0$ bosons, the neutral CP-odd $A^0$ boson,
and the charged $H^\pm$-boson pair.
The 2HDM has six free parameters, which are usually taken to be the masses
$m_{h^0}$, $m_{H^0}$, $m_{A^0}$, and $m_{H^\pm}$, the ratio
$\tan\beta=v_2/v_1$ of the vacuum expectation values of the two Higgs
doublets, and the weak mixing angle $\alpha$ that relates the weak and mass
eigenstates of $h^0$ and $H^0$.
At the tree level, the MSSM Higgs sector has just two free parameters, which
are usually taken to be the $m_{A^0}$ and $\tan\beta$.

The discovery of the $H^\pm$ bosons would prove the SM wrong and, at the same
time, give strong support to the 2HDM and the MSSM.
If the $H^\pm$ bosons have mass $m_H<m_t-m_b$, they will be mainly produced
through the $t\to bH^+$ decays of top quarks, which are copiously generated
singly or in pairs at an $e^+e^-$ LC \cite{kom}.
On the other hand, if there is sufficient center-of-mass (c.m.) energy
available, $\sqrt s>2m_H$, then charged-Higgs-boson pair production,
$e^+e^-\to H^+H^-$, will be the dominant production mechanism \cite{kom}.
Otherwise, if $m_H>\max(m_t-m_b,\sqrt s/2)$, the $H^\pm$ bosons can still be
produced singly.
There are various mechanisms of single-charged-Higgs-boson production
\cite{kan}.
The most important of them are $e^+e^-\to W^+H^-$, which proceeds through
quantum loops involving SM \cite{shi,arh} and possibly supersymmetric
\cite{shi} particles, $e^+e^-\to\tau^+\nu_\tau H^-$ \cite{kan,mor}, and
$e^+e^-\to t\overline{b}H^-$ \cite{kan,djo}.

\begin{figure}[ht]
\begin{center}
\leavevmode
\epsfxsize=14.cm
\epsffile[75 245 525 515]{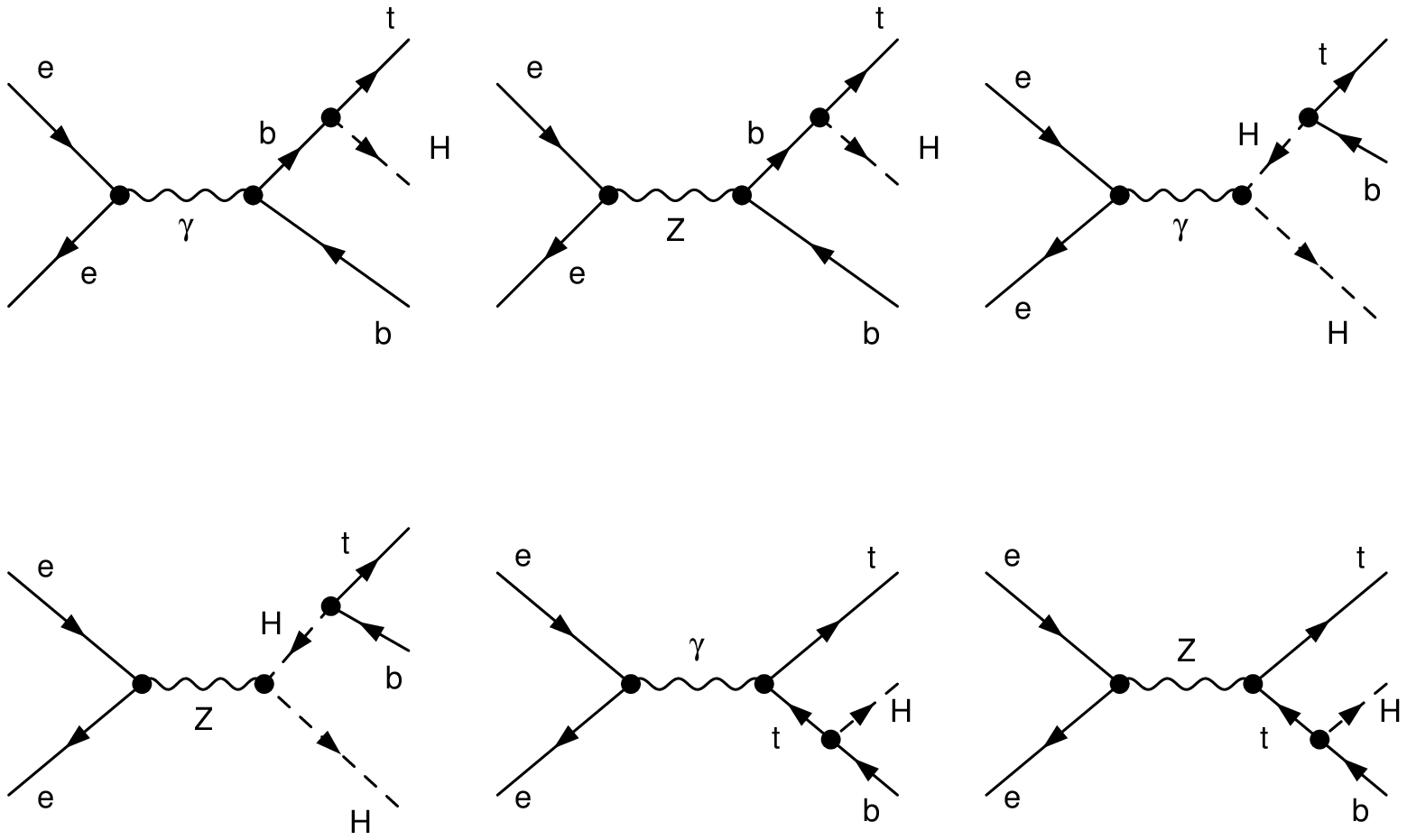}
\caption[]{\label{fig:born}Tree-level Feynman diagrams pertinent to the
process $e^+ e^-\to t\overline{b}H^-$.}
\end{center}
\end{figure}

In the following, we are concerned with the latter process.
At the tree level, it proceeds through the Feynman diagrams depicted in
Fig.~\ref{fig:born}.
It is kinematically allowed if $\sqrt s>m_t+m_b+m_H$.
Hence, we are most interested in a situation where
$m_t+m_b<\sqrt s/2<m_H<\sqrt s-m_t-m_b$.
For $\sqrt s=500$~GeV (800~GeV), this implies that $250\lsim m_H\lsim320$~GeV
($400\lsim m_H\lsim620$~GeV).
In such a situation, none of the virtual particles appearing in
Fig.~\ref{fig:born} can be resonating.
We note in passing that the absence of resonances is also guaranteed if
$m_t-m_b<m_H<m_t+m_b$ \cite{djo}.
However, this process is then of minor interest because we always have
$\sqrt s>2m_H$, so that $e^+e^-\to H^+H^-$ will take place.
In the presence of a resonance, the cross section approximately factorizes. 
Specifically, we have
$\sigma(e^+e^-\to t\overline{b}H^-)\approx\sigma(e^+e^-\to t\overline{t})
B(\overline{t}\to\overline{b}H^-)$ when the virtual top quark gets on its mass
shell, while we have
$\sigma(e^+e^-\to t\overline{b}H^-)\approx\sigma(e^+e^-\to H^+H^-)
B(H^+\to t\overline{b})$ when the virtual $H^+$ boson gets on its mass shell.

The purpose of this paper is to investigate the dominant quantum corrections
to the cross section of $e^+e^-\to t\overline{b}H^-$, which arise from quantum
chromodynamics (QCD).
The relevant Feynman diagrams emerge by attaching one gluon line in all
possible ways to each of the diagrams shown in Fig.~\ref{fig:born}.
This leads to $2\to3$ diagrams with one closed loop (see Fig.~\ref{fig:virt}),
which yield the virtual corrections, and to $2\to4$ diagrams of the tree-level
type (see Fig.~\ref{fig:real}), which give rise to the real corrections.
The loop diagrams involve two, three, or four virtual particles.

While the QCD corrections to the cross sections of the related processes
$e^+e^-\to q\overline{q}\Phi$, where $q=t,b$ and $\Phi=H$ \cite{dit,daw} or
$\Phi=h^0,H^0,A^0$ \cite{rei}, are available in the literature, the
corresponding analysis for $e^+e^-\to t\overline{b}H^-$ has been lacking so
far.
The present paper fills this gap.

The cross section of the SM process $e^+e^-\to q\overline{q}H$ via a virtual
photon and its QCD corrections can be recovered from our results as a special
case, involving only a subclass of the Feynman diagrams shown in
Figs.~\ref{fig:born}--\ref{fig:real}.
As a by-product of our analysis, we confirm the numerical results for this
cross section obtained in Refs.~\cite{dit,daw}.
We also perform the complete calculation for this process, which involves
Feynman diagrams where the $H$ boson is radiated off a virtual $Z$-boson line,
and find good agreement with Ref.~\cite{dit}.

This paper is organized as follows.
In Sec.~\ref{sec:ana}, we list a compact Born formula for the cross section of
$e^+e^-\to t\overline{b}H^-$ and give details of our analytical calculation of
its virtual and real QCD corrections.
Lengthy expressions are relegated to Appendices~\ref{app:tree} and
\ref{app:four}, where the Born form factors and the parameterization of the
four-particle phase space, respectively, may be found.
In Sec.~\ref{sec:num}, we present our numerical results.
In Sec.~\ref{sec:con}, we conclude with a summary of our analysis.

\section{\label{sec:ana}Analytical results}

In this section, we list a compact Born formula for the cross section of
$e^+e^-\to t\overline{b}H^-$ and give details of our analytical calculation of
its virtual and real QCD corrections.
By charge-conjugation invariance, the results for
$e^+e^-\to \overline{t}bH^+$ are the same.

\subsection{Born cross section}

We start by defining the kinematics.
We call the four-momenta of the incoming electron and positron $k_1$ and $k_2$
and those of the outgoing $t$ quark, $\overline{b}$ quark, and $H^-$ boson
$p_1$, $p_2$, and $p_3$, respectively.
We neglect the electron mass, but retain the $b$-quark mass, so that the
on-mass-shell (OS) conditions read $k_1^2=k_2^2=0$, $p_1^2=m_t^2$,
$p_2^2=m_b^2$, and $p_3^2=m_H^2$.
The virtual photon and $Z$ boson have four-momentum $p=k_1+k_2=p_1+p_2+p_3$,
and we define $s=p^2$.
It is convenient to introduce the dimensionless Lorentz scalars $a_i=p_i^2/s$,
$x_i=2p\cdot p_i/s$, and $y_i=\sqrt{x_i^2-4a_i}$ ($i=1,2,3$).
In the c.m.\ frame, $x_i=2p_i^0/\sqrt s$ and
$y_i=2|\mbox{\boldmath$p_i$}|/\sqrt s$ carry the meaning of scaled energies
and absolute three-momenta, respectively.
By four-momentum conservation, we have $x_1+x_2+x_3=2$.

The differential Born cross section may be evaluated as
\begin{equation}
d\sigma_{\rm Born}=\frac{1}{2s}\,\frac{1}{4}|{\cal T}_{\rm Born}|^2
d{\rm PS}_3(p;p_1,p_2,p_3),
\label{eq:tree}
\end{equation}
where the first and second factors on the right-hand side stem from the flux
and the average over the lepton spins, respectively, ${\cal T}_{\rm Born}$ is
the transition-matrix element corresponding to the Feynman diagrams of
Fig.~\ref{fig:born}, and the summation over the lepton and quark spins is
implied.
We assume that the incoming leptons are unpolarized.
Here and in the following, we define the Lorentz-invariant $n$-particle
phase-space measure as
\begin{equation}
d{\rm PS}_n(p;p_1,\ldots,p_n)
=(2\pi)^4\delta^{(4)}\left(p-\sum_{i=1}^np_i\right)\prod_{i=1}^n
\frac{d^3p_i}{(2\pi)^32p_i^0}.
\end{equation}

We now discuss the parameterization of the three-particle phase space.
We wish to express the Born cross section differential with respect to the
scaled energies of the final-state quarks, $x_1$ and $x_2$.
For convenience, we work in the c.m.\ frame, define the $z$ axis of the
coordinate system to point along
$\mbox{\boldmath$k_1$}=-\mbox{\boldmath$k_2$}$, and fix the $x$ axis in an
arbitrary way.
We then have
\begin{equation}
d{\rm PS}_3(p;p_1,p_2,p_3)
=\frac{4}{(4\pi)^5}dp_1^0\,d\cos\theta_1\,d\phi_1\,dp_2^0\,d\phi_2,
\end{equation}
where $\theta_1$ and $\phi_1$ are the polar and azimuthal angles of
$\mbox{\boldmath$p_1$}$, respectively, and $\phi_2$ is the azimuthal angle of
$\mbox{\boldmath$p_2$}$ with respect to the axis pointing along
$\mbox{\boldmath$p_1$}$ measured from the plane spanned by
$\mbox{\boldmath$k_1$}$ and $\mbox{\boldmath$p_1$}$.
Due to the azimuthal symmetry of the problem at hand, the integration over
$\phi_1$ is trivial, and we may take $\mbox{\boldmath$p_1$}$ to lie in the
$x$-$z$ plane.
If we now rotate the coordinate system in such a way that
$\mbox{\boldmath$p_1$}$ points along the $z$ axis and $\mbox{\boldmath$p_2$}$
lies in the $x$-$z$ plane, then $\theta=\theta_1$ and $\phi=\pi-\phi_2$ define
the direction of $\mbox{\boldmath$k_1$}$.
We thus have
\begin{equation}
d{\rm PS}_3(p;p_1,p_2,p_3)=\frac{s}{2(4\pi)^4}dx_1\,dx_2\,d\cos\theta\,d\phi.
\end{equation}
Next, we observe that $|{\cal T}_{\rm Born}|^2$ can be written as a
contraction of two rank-two tensors, a leptonic one involving $k_1$ and $k_2$
and a quarkonic one involving $p_1$, $p_2$, and $p_3$.
The leptonic one has the form
\begin{equation}
L^{\mu\nu}=\tr\not k_1\gamma^\nu\left(v_e^\prime-a_e^\prime\gamma_5\right)
\not k_2\gamma^\mu(v_e-a_e\gamma_5),
\label{eq:lep}
\end{equation}
where $v_e$, $v_e^\prime$, $a_e$, and $a_e^\prime$ are generic vector and
axial-vector couplings of the electron to the photon or $Z$ boson.
Performing the integrations over $\cos\theta$ and $\phi$, we obtain
\begin{equation}
\int\frac{d\Omega}{4\pi}L^{\mu\nu}=\frac{4}{3}
\left(v_ev_e^\prime+a_ea_e^\prime\right)(p^\mu p^\nu-sg^{\mu\nu}).
\label{eq:int}
\end{equation}
The fact that Eq.~(\ref{eq:int}) just depends on $p$ dramatically simplifies
the remaining phase-space integrations, since scalar products of the type
$k_i\cdot p_j$ are precluded.
The residual scalar products are $p_i\cdot p_j=(s/2)(z_k-a_i-a_j)$, where
$z_k=1+a_k-x_k$, with $i,j,k=1,2,3$ and $i\ne j\ne k\ne i$.

We thus find the doubly-differential Born cross section to be
\begin{equation}
\frac{d\sigma_{\rm Born}}{dx_1\,dx_2}=\frac{G_F^3m_Z^4}{32\pi^3\sqrt 2}
\left[{\cal Q}_e^2f_{\gamma\gamma}(x_1,x_2)
+{\cal Q}_e{\cal V}_ef_{\gamma Z}(x_1,x_2)
+\left({\cal V}_e^2+{\cal A}_e^2\right)f_{ZZ}(x_1,x_2)\right],
\label{eq:born}
\end{equation}
where $G_F$ is Fermi's constant,
\begin{equation}
{\cal Q}_e=-2c_ws_wQ_e,\qquad
{\cal V}_e=\frac{I_e-2s_w^2Q_e}{1-m_Z^2/s},\qquad
{\cal A}_e=\frac{I_e}{1-m_Z^2/s},
\end{equation}
and $f_{\gamma\gamma}(x_1,x_2)$, $f_{\gamma Z}(x_1,x_2)$, and
$f_{ZZ}(x_1,x_2)$ are form factors listed in Appendix~\ref{app:tree}.
Here, $s_w^2=1-c_w^2=1-m_W^2/m_Z^2$ is the sine square of the weak mixing
angle, $m_W$ and $m_Z$ are the masses of the $W$ and $Z$ bosons, respectively,
$Q_e=-1$ is the electric charge of the electron, and $I_e=-1/2$ is the third
component of weak isospin of its left-handed component.

The boundaries of integration are
\begin{eqnarray}
2\sqrt{a_1}&<&x_1<1+a_1-\left(\sqrt{a_2}+\sqrt{a_3}\right)^2,
\nonumber\\
x_2^-&<&x_2<x_2^+,
\end{eqnarray}
with
\begin{equation}
x_2^\pm=\frac{1}{2z_1}\left[(2-x_1)(z_1+a_2-a_3)
\pm y_1\sqrt{\lambda(z_1,a_2,a_3)}\right],
\end{equation}
where $\lambda(x,y,z)=x^2+y^2+z^2-2(xy+yz+zx)$ is K\"all\'en's function.
We perform the integrations over $x_1$ and $x_2$ numerically with the aid of
the multi-dimensional Monte Carlo integration routine {\tt VEGAS}
\cite{vegas}.

\subsection{Virtual corrections}

\begin{figure}[ht]
\begin{center}
\leavevmode
\epsfxsize=14.cm
\epsffile[100 325 505 730]{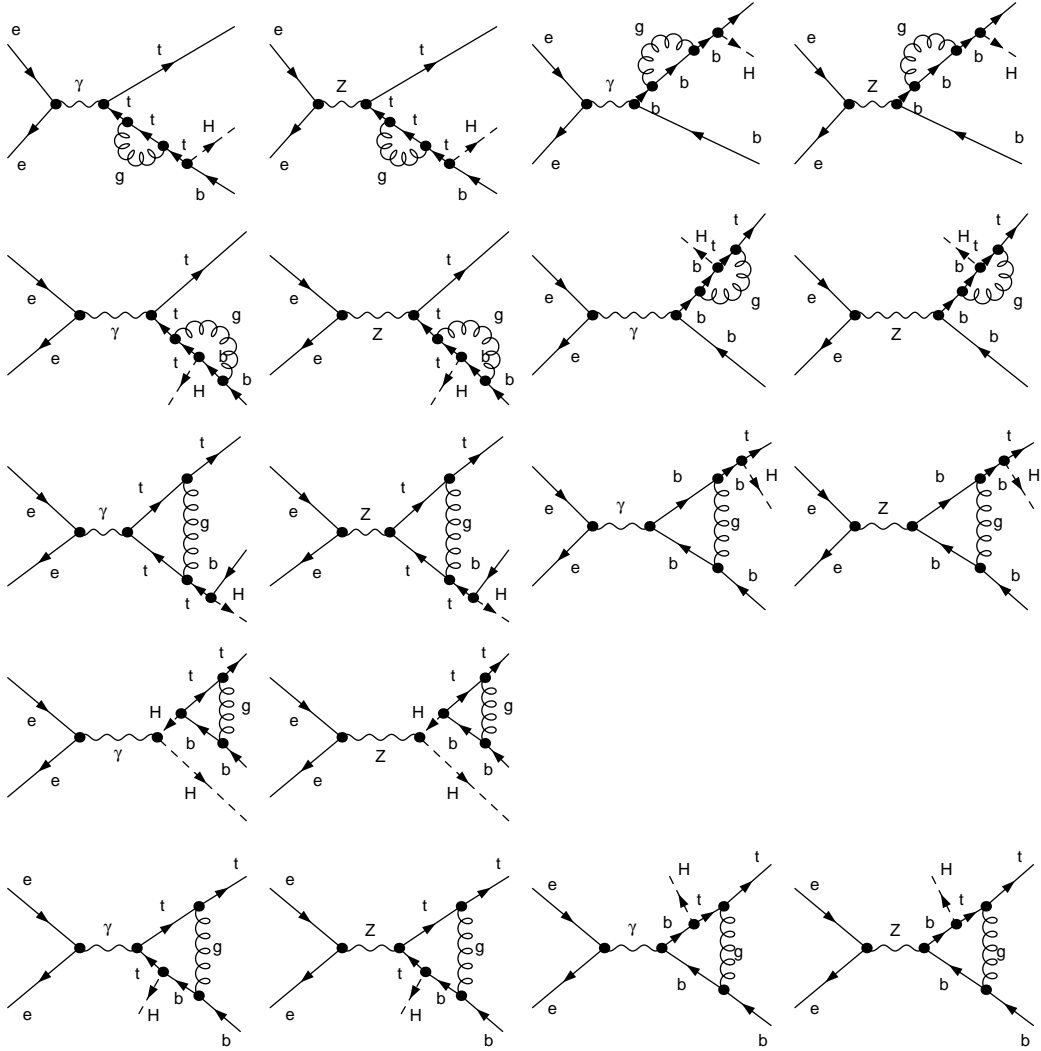}
\caption[]{\label{fig:virt}One-loop Feynman diagrams pertinent to the process
$e^+ e^-\to t\overline{b}H^-$.
They are classified according to the number of propagators in the loop.}
\end{center}
\end{figure}

We now turn to the virtual QCD corrections, which arise from the one-loop
Feynman diagrams shown in Fig.~\ref{fig:virt}.
Specifically, they include $t$- and $b$-quark self-energy corrections;
$t\overline{t}\gamma$, $t\overline{t}Z$, $b\overline{b}\gamma$,
$b\overline{b}Z$, and $t\overline{b}H^-$ vertex corrections; and
$t\overline{b}\gamma H^-$ and $t\overline{b}ZH^-$ box corrections.
These corrections suffer both from infrared (IR) and ultraviolet (UV)
divergences.
We regularize the former by endowing the gluon with an infinitesimal mass,
$m_g$.
In our case, this does not spoil gauge invariance, since the non-abelian
nature of QCD does not yet emerge at next-to-leading order (NLO).
This leads to terms logarithmic in $m_g$, which combine with similar terms
arising from soft-gluon emission, to be discussed below, to give a
$m_g$-independent result.
We establish this cancellation analytically.

UV divergences only occur in the self-energy and vertex corrections.
We extract them using dimensional regularization, with $D=4-2\epsilon$
space-time dimensions and 't~Hooft mass scale $\mu$.
They are removed by renormalization.
Specifically, we need to renormalize the quark masses and wave functions
appearing in ${\cal T}_{\rm Born}$.
Notice that the quark masses enter not only through the quark propagators, but
also through the $t\overline{b}H^-$ coupling.
To this end, we substitute $m_q\to m_q^0=m_q+\delta m_q$ and
$\psi_q\to\psi_q^0=\psi_q\sqrt{1+\delta Z_2^q}$ ($q=t,b$), where bare
quantities are denoted by the superscript 0.
In the OS scheme, the renormalization constants read
\begin{eqnarray}
\delta m_q&=&-\frac{\alpha_s}{4\pi}C_F\left[3\Delta+3\ln\frac{\mu^2}{m_q^2}
+4+{\cal O}(\epsilon)\right],
\label{eq:dm}\\
\delta Z_2^q&=&-\frac{\alpha_s}{4\pi}C_F\left[\Delta+\ln\frac{\mu^2}{m_q^2}
+2\ln\frac{m_g^2}{m_q^2}+4+{\cal O}(\epsilon)\right],
\end{eqnarray}
with $C_F=(N_c^2-1)/(2N_c)=4/3$ for $N_c=3$ quark colors and
\begin{equation}
\Delta=\frac{1}{\epsilon}-\gamma_E+\ln(4\pi),
\end{equation}
where $\gamma_E$ is the Euler-Mascheroni constant.
Notice that $Z_2^q$ is also IR divergent.
The expression for $\delta m_q$ in the modified minimal-subtraction
($\overline{\rm MS}$) scheme \cite{bar} emerges from Eq.~(\ref{eq:dm}) by
retaining only the first term contained within the square brackets.

The virtual QCD corrections may be evaluated as
\begin{equation}
d\sigma_{\rm virt}=d\sigma_{\rm Born}\delta_{\rm virt}(m_g),
\end{equation}
with
\begin{equation}
\delta_{\rm virt}(m_g)=\frac{2}{|{\cal T}_{\rm Born}|^2}
\re\left\{{\cal T}_{\rm Born}^*\left[\sum_{q=t,b}\left(\delta m_q
\frac{\partial}{\partial m_q}+\frac{1}{2}\delta Z_2^q\right)
{\cal T}_{\rm Born}+{\cal T}_{\rm loop}\right]\right\},
\label{eq:virt}
\end{equation}
where ${\cal T}_{\rm loop}$ is the transition-matrix element corresponding to
the Feynman diagrams of Fig~\ref{fig:virt}.
Notice that the quark masses that appear in the squares of the quark spinors
and in the boundaries of the phase-space integration correspond to
renormalized ones from the outset.
As mentioned above, $\delta_{\rm virt}(m_g)$ is UV finite, but IR divergent. 
Notice that Eq.~(\ref{eq:int}), which refers to the physical case $D=4$, can
still be used at the one-loop level, since the quarkonic tensor is by itself
UV finite upon renormalization. 

We generate ${\cal T}_{\rm loop}$ and reduce it to standard one-loop scalar
integrals in two independent ways:
one is based on the combination of the program packages {\tt FeynArts}
\cite{feynarts} and {\tt FormCalc} \cite{formcalc} and the other one on
custom-made routines written in the program language {\tt FORM} \cite{form}.
We then evaluate the standard one-loop scalar integrals, the IR-divergent ones
analytically using the results of Ref.~\cite{bee} and the IR-finite ones
numerically with the help of the program package {\tt LoopTools}
\cite{formcalc}.
Our analytic result for $\delta_{\rm virt}(m_g)$ is too lengthy to be
presented here.

\subsection{Real corrections}

\begin{figure}[ht]
\begin{center}
\leavevmode
\epsfxsize=14.cm
\epsffile[70 155 540 605]{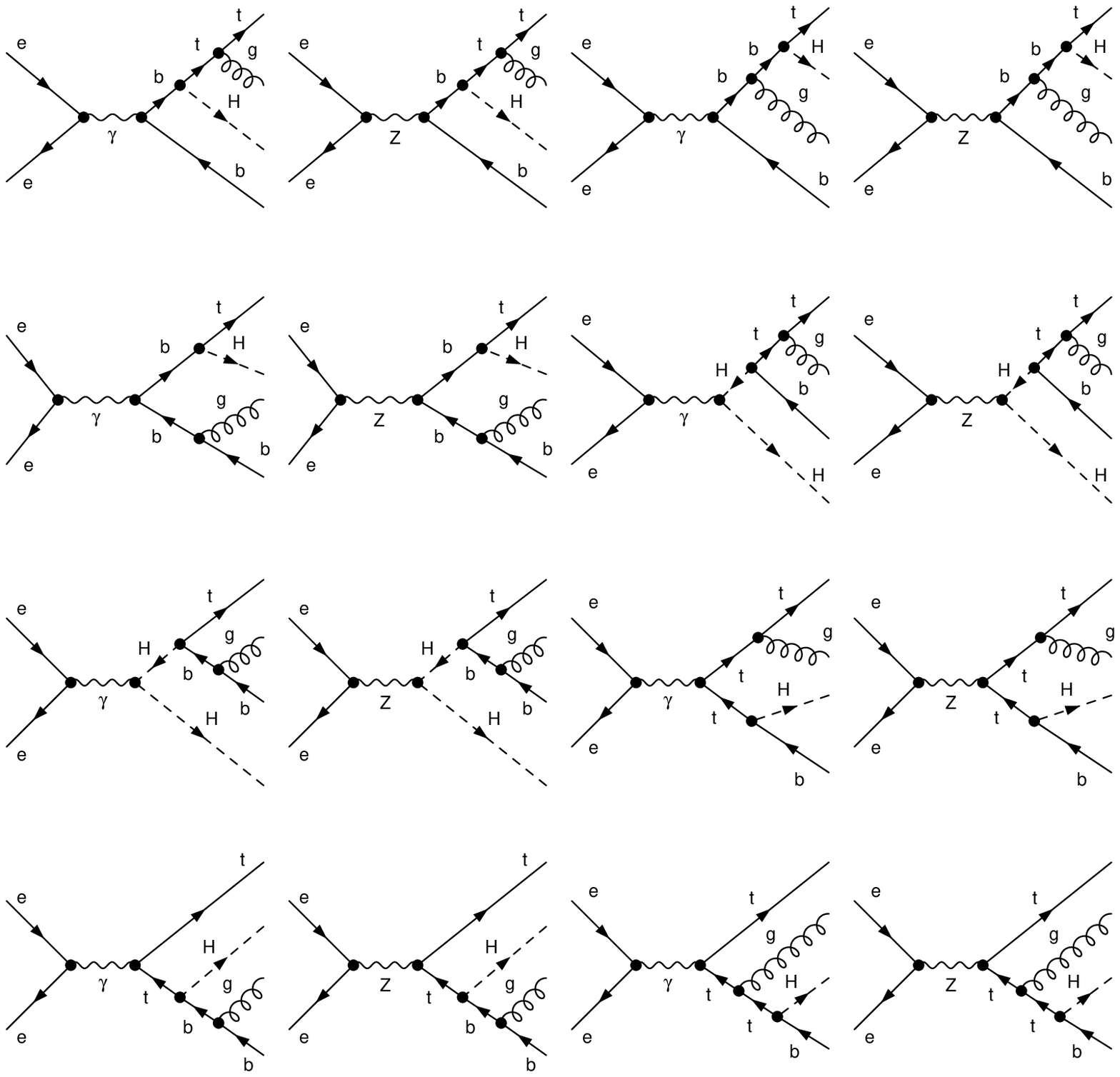}
\caption[]{\label{fig:real}Tree-level Feynman diagrams pertinent to the
process $e^+ e^-\to t\overline{b}gH^-$.}
\end{center}
\end{figure}

We now proceed to the real QCD corrections, which arise from the $2\to4$
tree-level Feynman diagrams shown in Fig.~\ref{fig:real}.
We denote the gluon four-momentum by $q$.
As mentioned above, the emission of soft gluons generates IR divergences.
For consistency with the evaluation of the virtual QCD corrections, the latter
must also be regularized by the gluon mass $m_g$.
It is convenient to work in the c.m.\ frame and to introduce an unphysical
gluon-energy cutoff, $E_{\rm cut}$, with $m_g\ll E_{\rm cut}\ll E_{\rm max}$,
where $E_{\rm max}$ is the maximum gluon energy allowed by kinematics, so as
to separate the gluon phase space into soft and hard regions, defined by
$m_g<q^0<E_{\rm cut}$ and $E_{\rm cut}<q^0<E_{\rm max}$, respectively.
This has two technical advantages, since soft gluons with infinitesimal mass
$m_g$ do not affect the kinematics of the underlying process, while hard
gluons with zero mass do not produce IR divergences.
On the one hand, the soft-gluon bremsstrahlung may be treated analytically in
the eikonal approximation, which is independent of the underlying process and
results in a multiplicative correction to the Born result.
On the other hand, $m_g$ may be safely neglected in the treatment of the
hard-gluon bremsstrahlung, which facilitates the phase-space integration.
In turn, the soft- and hard-gluon contributions both depend on $E_{\rm cut}$,
while their combined contribution is, of course, independent of $E_{\rm cut}$,
which we checked numerically.
As mentioned above, the $m_g$ dependence of the soft-gluon contribution
analytically cancels against the one of the virtual QCD corrections.

The soft-gluon contribution is given by
\begin{equation}
d\sigma_{\rm soft}=d\sigma_{\rm Born}\delta_{\rm soft}(m_g,E_{\rm cut}),
\end{equation}
with
\begin{equation}
\delta_{\rm soft}(m_g,E_{\rm cut})
=-\frac{\alpha_sC_F}{(2\pi)^2}
\int_{|\mbox{\scriptsize\boldmath$q$}|<E_{\rm cut}}
\frac{d^3q}{q^0}\left(\frac{p_1}{p_1\cdot q}-\frac{p_2}{p_2\cdot q}\right)^2,
\label{eq:soft}
\end{equation}
where $q^0=\sqrt{\mbox{\boldmath$q$}^2+m_g^2}$ is the gluon energy.
The integration in Eq.~(\ref{eq:soft}) can be performed analytically as
described in Ref.~\cite{tho}, the result being
\begin{eqnarray}
\lefteqn{\delta_{\rm soft}(m_g,E_{\rm cut})=\frac{\alpha_s}{\pi}C_F\left\{
\left(\frac{\alpha}{2}\,\frac{z_3-a_1-a_2}{\alpha^2a_1-a_2}
\ln\frac{\alpha^2a_1}{a_2}-1\right)\ln\frac{4E_{\rm cut}^2}{m_g^2}\right.}
\nonumber\\
&&{}+\alpha\frac{z_3-a_1-a_2}{\alpha^2a_1-a_2}
\left[\li\left(1-\alpha\frac{x_1+y_1}{v}\right) 
+\li\left(1-\alpha\frac{x_1-y_1}{v}\right)\right.
\nonumber\\
&&{}-\left.\li\left(1-\frac{x_2+y_2}{v}\right) 
-\li\left(1-\frac{x_2-y_2}{v}\right)
+\frac{1}{4}\left(\ln^2\frac{x_1-y_1}{x_1+y_1}
-\ln^2\frac{x_2-y_2}{x_2+y_2}\right)\right]
\nonumber\\
&&{}-\left.\frac{1}{2}\left(\frac{x_1}{y_1}\ln\frac{x_1-y_1}{x_1+y_1} 
+\frac{x_2}{y_2}\ln\frac{x_2-y_2}{x_2+y_2}\right)\right\},
\label{eq:tho}
\end{eqnarray}
where $\li(x)=-\int_0^xdy\,\ln(1-y)/y$ is the Spence function,
\begin{eqnarray}
\alpha&=&\frac{1}{2a_1}\left[z_3-a_1-a_2+\sqrt{\lambda(z_3,a_1,a_2)}\right],
\nonumber\\
v&=&2\frac{\alpha^2a_1-a_2}{\alpha x_1-x_2}.
\end{eqnarray}
Notice that Eq.~(\ref{eq:tho}) is invariant under the interchange of the
indices 1 and 2.

The hard-gluon contribution may be evaluated by integrating
\begin{equation}
d\sigma_{\rm hard}(E_{\rm cut})=\frac{1}{2s}\,\frac{1}{4}
|{\cal T}_{\rm real}|^2\theta(q^0-E_{\rm cut})d{\rm PS}_4(p;p_1,p_2,p_3,q),
\label{eq:hard}
\end{equation}
where ${\cal T}_{\rm real}$ is the transition-matrix element corresponding to
the Feynman diagrams of Fig.~\ref{fig:real}, over the full four-particle phase
space, imposing the condition $q^0>E_{\rm cut}$.
We use the parameterization of the four-particle phase space presented in
Appendix~\ref{app:four}.
It involves five nontrivial integrations, which we perform numerically using
the Monte Carlo routine {\tt VEGAS} \cite{vegas}.
Our formula for $|{\cal T}_{\rm real}|^2$ is too lengthy to be listed here.

We performed several checks for our implementation of the four-particle
phase-space integration.
We numerically verified the analytical formula for the total cross section of
$e^+e^-\to q\overline{q}g^*\to q\overline{q}Q\overline{Q}$, where $q$ and $Q$
represent massless and massive quarks, respectively, and $g^*$ denotes a
virtual gluon, given in Eq.~(2) of Ref.~\cite{hoa}.
In this case, IR singularities do not appear in intermediate steps, so that no
separation into soft-gluon and hard-gluon contributions is required.
We also found excellent agreement with a numerical result for a similar
process involving four different quark masses obtained using the democratic
multi-particle phase-space generator {\tt RAMBO} \cite{rambo}.

Our final result for the QCD-corrected differential cross section reads
\begin{equation}
d\sigma_{\rm QCD}=d\sigma_{\rm Born}[1+\delta_{\rm virt}(m_g)
+\delta_{\rm soft}(m_g,E_{\rm cut})]+d\sigma_{\rm hard}(E_{\rm cut}),
\end{equation}
where $d\sigma_{\rm Born}$, $\delta_{\rm virt}(m_g)$,
$\delta_{\rm soft}(m_g,E_{\rm cut})$, and $d\sigma_{\rm hard}(E_{\rm cut})$
are defined in Eqs.~(\ref{eq:tree}), (\ref{eq:virt}), (\ref{eq:soft}), and
(\ref{eq:hard}), respectively.
It is manifestly independent of $m_g$ and insensitive to the choice of
$E_{\rm cut}$, as long as $m_g\ll E_{\rm cut}\ll E_{\rm max}$, as we verified
numerically.
We also checked that the QCD-corrected total cross section is finite in the
limit $m_b\to0$, in compliance with the Kinoshita-Lee-Nauenberg theorem
\cite{kln}.

The QCD-corrected cross section of the SM process $e^+e^-\to q\overline{q}H$
via a virtual photon can be obtained from our results as a special case,
involving only a subclass of the Feynman diagrams shown in
Figs.~\ref{fig:born}--\ref{fig:real}.
As a by-product of our analysis, we confirmed the numerical results for this
cross section obtained in Refs.~\cite{dit,daw}.
We also performed the complete calculation for this process, which involves
Feynman diagrams where the $H$ boson is radiated off a virtual $Z$-boson line,
and found good agreement with Ref.~\cite{dit}.
In turn, this provides a nontrivial check for all parts of our analysis.

\section{\label{sec:num}Numerical results}

We are now in a position to present our numerical results.
We first specify our input parameters.
We use $m_W=80.419$~GeV, $m_Z=91.1882$~GeV, $m_t=174.3$~GeV, $m_b=4.6$~GeV,
$G_F=1.16639\times10^{-5}$~GeV${}^{-2}$ \cite{pdg}, and the present world
average $\alpha_s^{(5)}(m_Z)=0.1180$ \cite{kkp}.
We consistently evaluate $\alpha_s^{(n_f)}(\mu)$ and
$\overline{m}_q^{(n_f)}(\mu)$ to lowest order (LO) in the $\overline{\rm MS}$
scheme with $n_f=6$ active quark flavors performing the matching with $n_f=5$
QCD at scale $m_t$.
For the reader's convenience, we collect the relevant formulas here
\cite{kni}:
\begin{eqnarray}
\frac{1}{\alpha_s^{(5)}(\mu)}&=&\frac{1}{\alpha_s^{(5)}(m_Z)}
+\frac{\beta_0^{(5)}}{\pi}\ln\frac{\mu^2}{m_Z^2},
\\
\frac{1}{\alpha_s^{(6)}(\mu)}&=&\frac{1}{\alpha_s^{(5)}(m_t)}
+\frac{\beta_0^{(6)}}{\pi}\ln\frac{\mu^2}{m_t^2},
\\
\overline{m}_t^{(6)}(\mu)&=&m_t\left[1-\frac{\alpha_s^{(6)}(m_t)}{\pi}C_F
\right]\left[\frac{\alpha_s^{(6)}(\mu)}{\alpha_s^{(6)}(m_t)}\right]
^{\gamma_0/\beta_0^{(6)}},
\label{eq:mt}\\
\overline{m}_b^{(6)}(\mu)&=&m_b\left[1-\frac{\alpha_s^{(5)}(m_b)}{\pi}C_F
\right]\left[\frac{\alpha_s^{(5)}(m_t)}{\alpha_s^{(5)}(m_b)}\right]
^{\gamma_0/\beta_0^{(5)}}
\left[\frac{\alpha_s^{(6)}(\mu)}{\alpha_s^{(6)}(m_t)}\right]
^{\gamma_0/\beta_0^{(6)}},
\end{eqnarray}
where
\begin{eqnarray}
\beta_0^{(n_f)}&=&{1\over4}\left({11\over3}C_A-{4\over3}T_Fn_f\right),
\nonumber\\
\gamma_0&=&{3\over4}C_F,
\end{eqnarray}
with $C_A=N_c$ and $T_F=1/2$, are the first coefficients of the
Callan-Symanzik beta function and the quark-mass anomalous dimension,
respectively.
For simplicity, we use a common renormalization scale $\mu$ in
$\alpha_s^{(6)}(\mu)$ and $\overline{m}_q^{(6)}(\mu)$.
We study the cases $\sqrt s=500$~GeV and 800~GeV.
As for the MSSM input parameters, we consider the ranges
$1<\tan\beta<40\approx m_t/m_b$ and $250<m_H<320$~GeV if $\sqrt s=500$~GeV or
$400<m_H<620$~GeV if $\sqrt s=800$~GeV.

We now discuss the influence of the QCD corrections on the total cross
sections of $e^+e^-\to t\overline{b}H^-$ and its charge conjugate counterpart,
which we add.  
We start by selecting the renormalization scheme and scale that are most
appropriate for the problem under consideration.
For this purpose, we study the $\mu$ dependence of the Born and QCD-corrected
results in two different renormalization schemes.
The first one uses the pole masses $m_t$ and $m_b$ as basic parameters (OS
scheme), while the second one uses $m_t$ and the $\overline{\rm MS}$ mass
$\overline{m}_b^{(6)}(\mu)$ as basic parameters (mixed scheme).
Both schemes employ the $\overline{\rm MS}$ definition of
$\alpha_s^{(6)}(\mu)$.
We refrain from utilizing $\overline{m}_t^{(6)}(\mu)$, which, in general,
significantly deviates from $m_t$, as may be seen from Eq.~(\ref{eq:mt}).
For, if we were to include the weak decays of the $t$ and $\overline{t}$
quarks in our analysis, then, during the propagation of these quarks between
their production and decay vertices, configurations near their {\it physical}
mass shells would be kinematically favored.
As a matter of fact, the experimentally measured invariant masses of their
decay products are very close to $m_t$.
In turn, the phase space of $e^+e^-\to t\overline{b}H^-$ would undergo a
significant, yet artificial change of size if it were parameterized in terms
of $\overline{m}_t^{(6)}(\mu)$ rather than $m_t$.
On the other hand, the use of $\overline{m}_b^{(6)}(\mu)$ is predicated on the
grounds that it automatically resums large logarithmic corrections that arise
if the $t\overline{b}H^-$ Yukawa coupling is expressed in terms of $m_b$.
A similar feature is familiar from the $H\to b\bar b$ decay in the SM
\cite{bra}.
This effect is particularly pronounced for large values of $\tan\beta$ because
the $t\overline{b}H^-$ Yukawa coupling is then approximately proportional to
the $b$-quark mass.

\begin{figure}[ht]
\begin{center}
\leavevmode
\epsfxsize=14.cm
\epsffile[45 240 550 470]{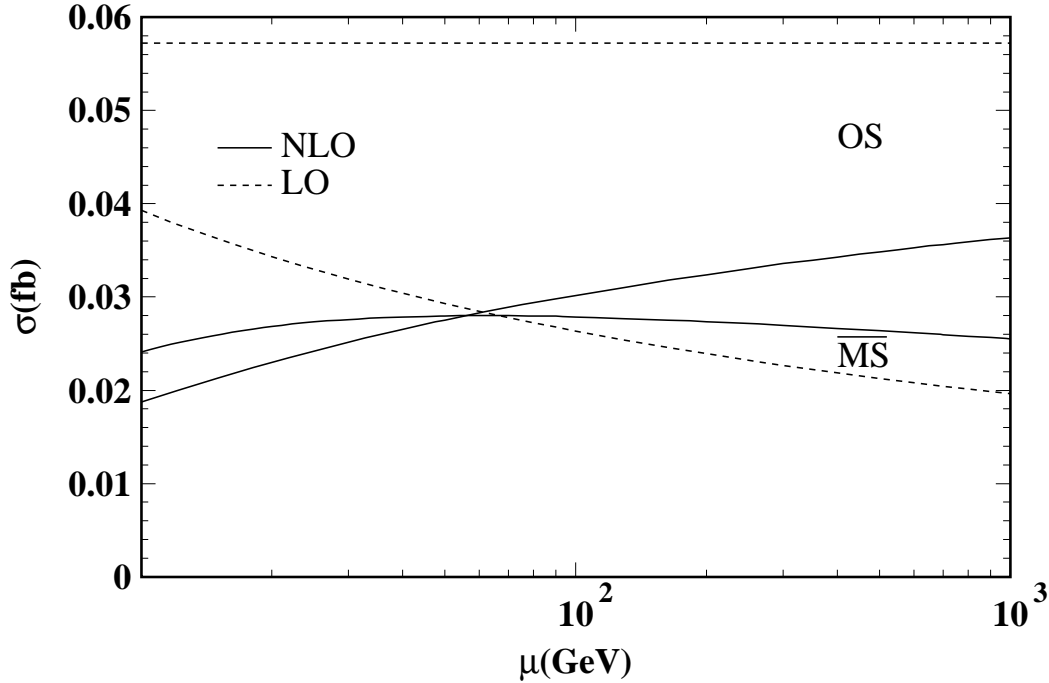}
\caption[]{\label{fig:scheme}
Total cross section of $e^+e^-\to t\overline{b}H^-,\overline{t}bH^+$ as a
function of $\mu$ for $\sqrt s=500$~GeV, $\tan\beta=40$, and $m_H=260$~GeV.
The dashed and solid curves correspond to the Born and QCD-corrected results,
respectively.
The upper and lower sets of curves refer to the OS and mixed schemes,
respectively.}
\end{center}
\end{figure}

As a typical example, we consider in Fig.~\ref{fig:scheme} the case of
$\sqrt s=500$~GeV, $\tan\beta=40$, and $m_H=260$~GeV.
We allow $\mu$ to vary over two orders of magnitude, from 10 to 1000~GeV.
In the OS scheme, the Born result is $\mu$ independent, while the
QCD-corrected one depends on $\mu$ via $\alpha_s^{(6)}(\mu)$.
In the mixed scheme, the $\mu$ dependence enters at LO via
$\overline{m}_b^{(6)}(\mu)$ and at NLO via $\alpha_s^{(6)}(\mu)$ and
$\overline{m}_b^{(6)}(\mu)$.
Obviously, the theoretical uncertainties due to scheme and typical scale
variations are significantly reduced as we pass from LO to NLO.
On the one hand, the OS-scheme to mixed-scheme ratio is brought down to the
vicinity of unity, from 1.46--2.92 to 0.78--1.43, depending on the value of
$\mu$.
On the other hand, the $\mu$ dependence within the mixed scheme is reduced by
a factor of 5, from 0.020~fb to 0.004~fb in absolute terms.
Furthermore, we observe that, in the OS scheme, the QCD corrections lead to a
dramatic reduction of the cross section, by 36--67\%.
As explained above, this is because they contain large logarithmic terms of
the form $\alpha_s^{(6)}(\mu)\ln\left(M^2/m_b^2\right)$, where $M$ is a
generic mass scale in the ball park of some suitable average of the
final-state-particle masses, $m_b$, $m_t$, and $m_H$.
In the mixed scheme with $\mu$ of order $M$, such terms are shifted from the
QCD corrections to the Born result, where they are absorbed into the running
of $\overline{m}_b^{(6)}(\mu)$ from $\mu=m_b$ to $\mu=M$.
This is reflected in Fig.~\ref{fig:scheme} by the fact that, in the mixed
scheme, the QCD corrections are relatively modest, ranging from $-39\%$ to
$+30\%$.
Unless otherwise stated, we shall henceforth work in the mixed scheme, which,
for plausible values of $\mu$, is superior to the OS scheme as far as the
convergence properties are concerned.

Let us now turn to the question of how to fix the value of $\mu$ in a
reasonable way.
Scale-setting procedures frequently discussed in the literature include the
concept of fastest apparent convergence (FAC) \cite{fac}, the principle of
minimal sensitivity (PMS) \cite{pms}, and the proposal by Brodsky, Lepage, and
Mackenzie (BLM) \cite{blm} to resum the leading light-quark contribution to
the renormalization of the strong-coupling constant.
The latter is not yet applicable to the problem under consideration, which is
of LO in the strong-coupling constant.
The FAC and PMS prescriptions lead us to select the values of $\mu$ where the
Born and QCD-corrected results intersect and where the latter exhibits a local
extremum, respectively.
We observe from Fig.~\ref{fig:scheme} that these two $\mu$ values 
approximately coincide, at about 60~GeV.
Incidentally, in the close vicinity of these two $\mu$ values, also the
QCD-corrected results in the OS and mixed schemes cross over, so that also the
scheme dependence at NLO vanishes in this neighborhood, at least as for the
two schemes considered here.
Since the $\mu$ dependence is logarithmic, a democratic way of combining the
three scales $m_b$, $m_t$, and $m_H$ is by taking their geometric means,
$\mu=\sqrt[3]{m_bm_tm_H}$.
In the present case, this educated guess yields $\mu\approx60$~GeV, which
nicely agrees with the triply distinguished point identified above.
We checked that this choice works similarly well for the case of
$\sqrt s=800$~GeV, $\tan\beta=40$, and $m_H=410$~GeV.
We shall henceforth employ it, with the understanding that
Fig.~\ref{fig:scheme} provides us with a useful estimate of the theoretical
uncertainties due to scheme and typical scale variations, both at LO and NLO.

\begin{figure}[ht]
\begin{center}
\leavevmode
\epsfxsize=14.cm
\epsffile[55 240 540 570]{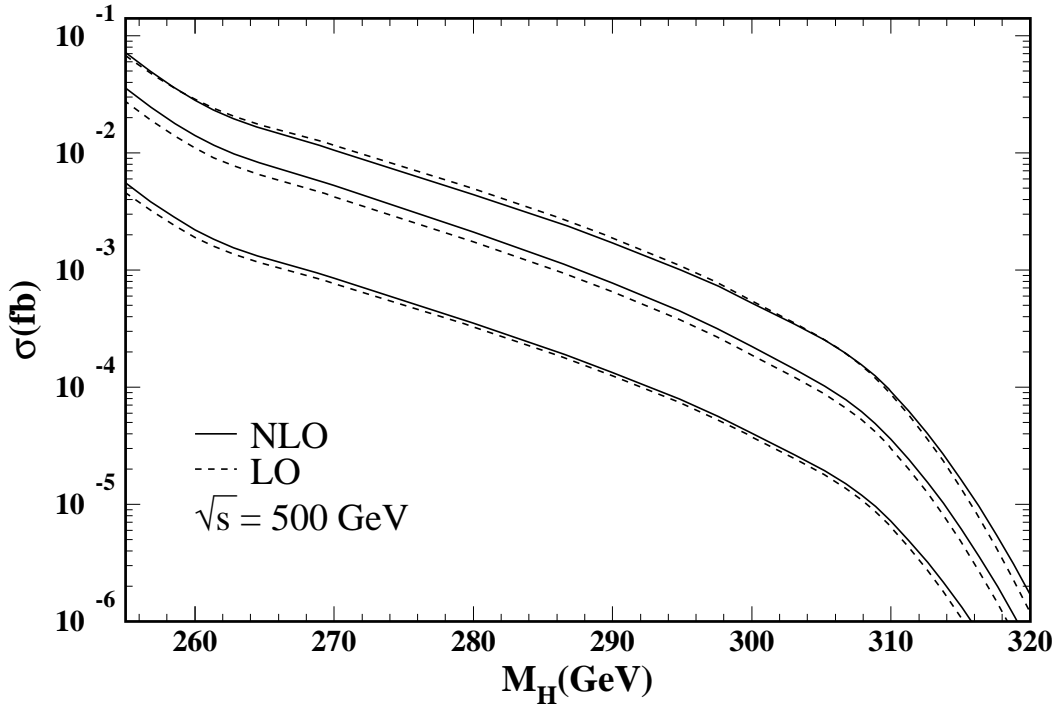}
\caption[]{\label{fig:mh_500}
Total cross section of $e^+e^-\to t\overline{b}H^-,\overline{t}bH^+$ without
(dotted curves) and with (solid curves) QCD corrections in the mixed scheme as
a function of $m_H$ for $\sqrt s=500$~GeV and various values of $\tan\beta$.
The middle, lower, and upper sets of curves correspond to $\tan\beta=2$, 6,
and 40, respectively.}
\end{center}
\end{figure}

\begin{figure}[ht]
\begin{center}
\leavevmode
\epsfxsize=14.cm
\epsffile[45 240 540 570]{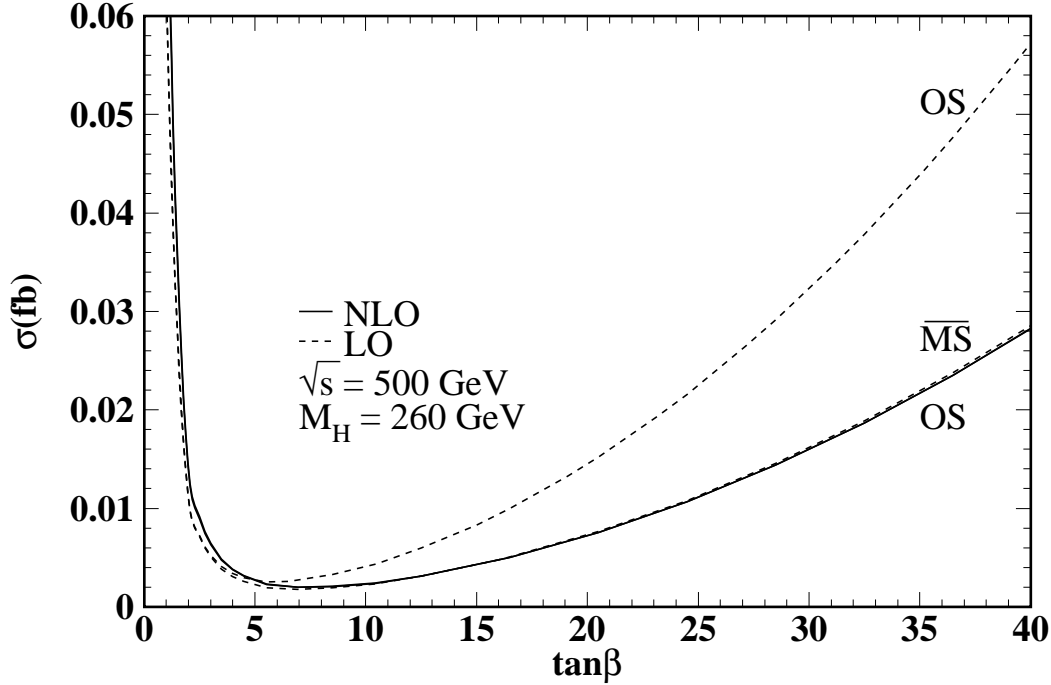}
\caption[]{\label{fig:tb_500}Total cross section of
$e^+e^-\to t\overline{b}H^-,\overline{t}bH^+$ without (dotted curves) and with
(solid curves) QCD corrections as a function of $\tan\beta$ for
$\sqrt s=500$~GeV and $m_H=260$~GeV.
The upper and lower dotted curves refer to the OS and mixed schemes,
respectively.
The two solid curves referring to the OS and mixed schemes lie on top of each
other.}
\end{center}
\end{figure}

\begin{figure}[ht]
\begin{center}
\leavevmode
\epsfxsize=14.cm
\epsffile[55 240 545 565]{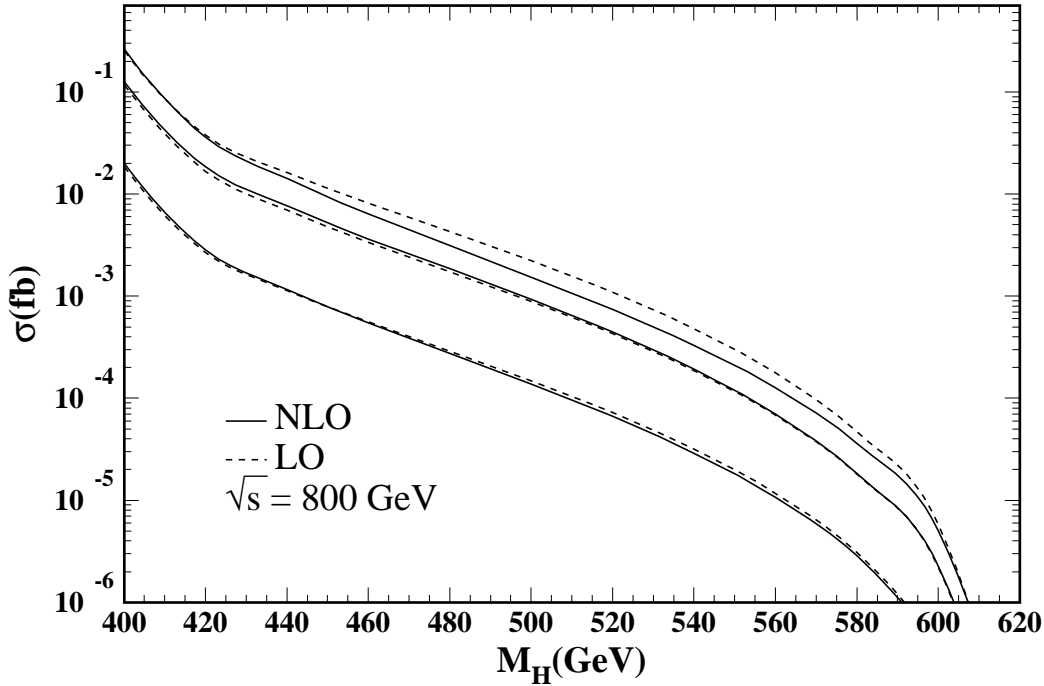}
\caption[]{\label{fig:mh_800}Same as in Fig.~\ref{fig:mh_500}, for
$\sqrt s=800$~GeV.}
\end{center}
\end{figure}

\begin{figure}[ht]
\begin{center}
\leavevmode
\epsfxsize=14.cm
\epsffile[50 240 540 565]{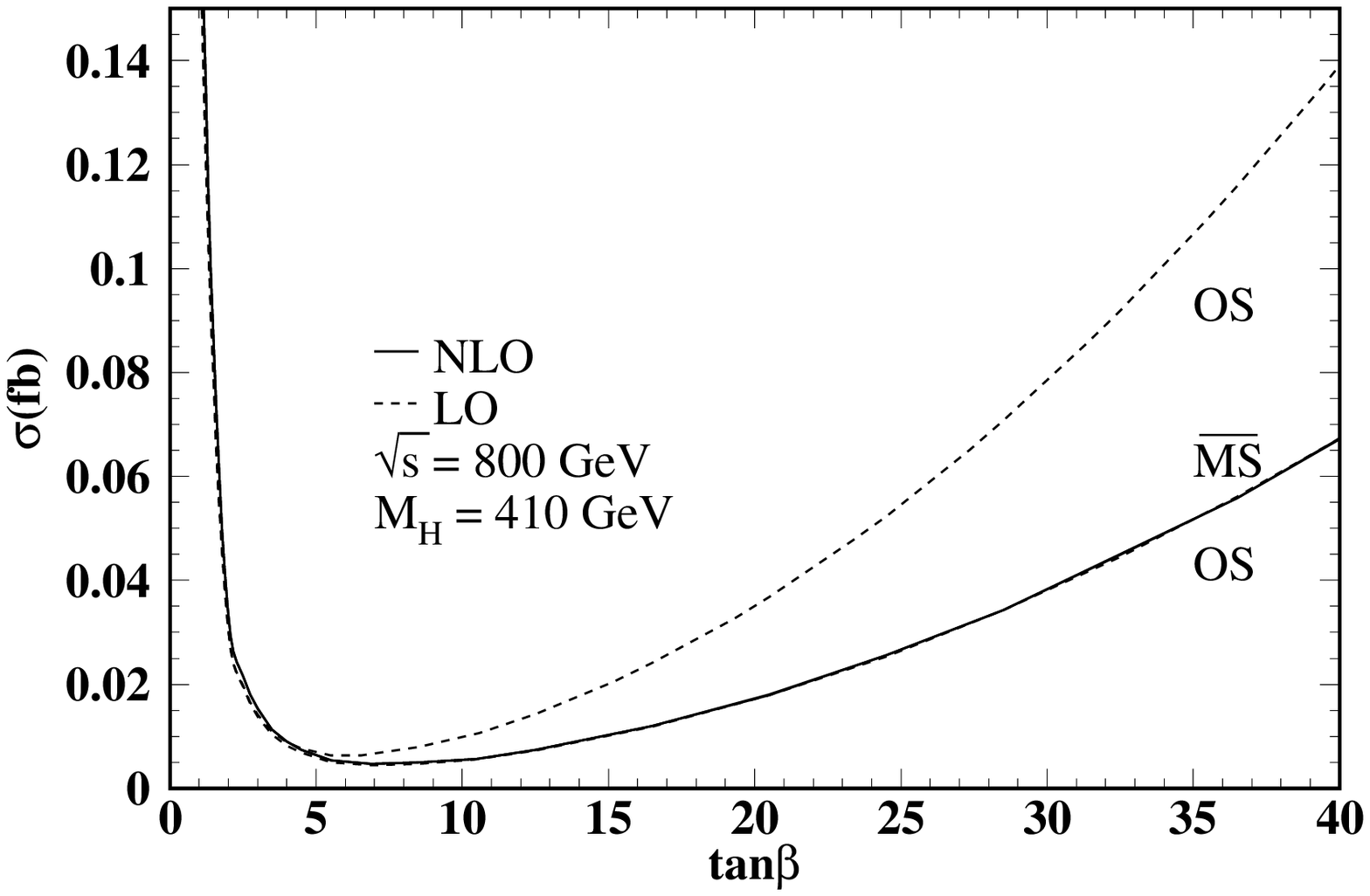}
\caption[]{\label{fig:tb_800}
Same as in Fig.~\ref{fig:tb_500}, for $\sqrt s=800$~GeV and $m_H=410$~GeV.}
\end{center}
\end{figure}

Figures~\ref{fig:mh_500} and \ref{fig:tb_500} refer to $\sqrt s=500$~GeV,
while Figs.~\ref{fig:mh_800} and \ref{fig:tb_800} refer to $\sqrt s=800$~GeV.
We investigate the $m_H$ dependence for various values of $\tan\beta$ in
Figs.~\ref{fig:mh_500} and \ref{fig:mh_800} and the $\tan\beta$ dependence
for typical values of $m_H$ in Figs.~\ref{fig:tb_500} and \ref{fig:tb_800}.
In each figure, we present the Born and QCD-corrected results in the mixed
scheme with $\mu=\sqrt[3]{m_bm_tm_H}$.
For comparison, in Figs.~\ref{fig:tb_500} and \ref{fig:tb_800}, we also
present the corresponding results in the OS scheme with the same scale choice.
We observe that the total cross sections exhibit minima close to
$\tan\beta\approx\sqrt{m_t/m_b}\approx6$, independently of order and scheme.
This may be understood by observing that the average strength of the
$t\overline{b}H^-$ coupling, which is proportional to
$\sqrt{m_t^2\cot^2\beta+m_b^2\tan^2\beta}$, is then minimal \cite{kra}.
Depending on $\sqrt s$, $\tan\beta$, and $m_H$, the QCD corrections may be of
either sign.
By construction, they are generally rather modest in the mixed scheme,
although they may reach a magnitude of 50\% for specific values of
$\sqrt s$, $\tan\beta$, and $m_H$, as may be seen from Fig.~\ref{fig:mh_800}.
On the other hand, in the OS scheme, the QCD corrections lead to a substantial
reduction in cross section at large values of $\tan\beta$, by up to 50\%.
As explained above, this may be attributed to large logarithms arising from
the $t\overline{b}H^-$ Yukawa coupling.
Finally, we notice that Figs.~\ref{fig:tb_500} and \ref{fig:tb_800} support
the observations made in connection with Fig.~\ref{fig:scheme}.
In fact, owing to our judicious scale choice, the Born and QCD-corrected
results in the mixed scheme and the QCD-corrected result in the OS scheme all
approximately coincide, which nicely demonstrates the perturbative stability
in the mixed scheme and the feeble scheme dependence at NLO.
By contrast, the perturbative stability in the OS scheme is rather poor at
large values of $\tan\beta$.

It is interesting to investigate the relative importance of the contributions
due to photon and $Z$-boson exchanges.
To this end, we evaluate the photon-induced part of the total cross section by
putting ${\cal V}_e={\cal A}_e=0$ and compare it with the full result.
We find that the bulk of the total cross section is due to photon exchange.
In fact, for the typical values $\sqrt s=500$~GeV and $m_H=260$~GeV, the
photon-induced part exhausts 78\%, 80\%, and 82\% of the full result if
$\tan\beta=2$, 6, and 40, respectively.

\section{\label{sec:con}Summary}

We considered the process $e^+e^-\to t\overline{b}H^-$ and its
charge-conjugate counterpart in the MSSM, which are among the dominant
charged-Higgs-boson production mechanisms at a future $e^+e^-$ LC if
$m_t+m_b<\sqrt s/2<m_H<\sqrt s-m_t-m_b$, so that $t\to bH^+$ and
$e^+e^-\to H^+H^-$ are not allowed kinematically.
For $\sqrt s=500$~GeV and 800~GeV, this corresponds to the $m_H$ windows
$250\lsim m_H\lsim320$~GeV and $400\lsim m_H\lsim620$~GeV, respectively.
We presented a compact Born formula for the cross sections of these processes
and evaluated their dominant radiative corrections, which arise from QCD.

We regularized the IR singularities by introducing an infinitesimal gluon mass
$m_g$ and the UV ones by using dimensional regularization.
The IR singularities cancelled when the virtual and soft real QCD corrections
were combined and the UV ones upon renormalizing the masses and wave functions
of the quarks in the Born transition-matrix element.
We established these cancellations analytically.
We separated the soft-gluon and hard-gluon contributions by introducing an
unphysical gluon-energy cutoff $E_{\rm cut}$ and verified that their sum is
insensitive to the choice of $E_{\rm cut}$ as long as
$m_g\ll E_{\rm cut}\ll E_{\rm max}$ is satisfied.

We worked in a mixed renormalization scheme, where the strong-coupling
constant and the $b$-quark mass are defined in the $\overline{\rm MS}$ scheme,
while the $t$-quark mass is defined in the OS scheme.
In this way, large logarithmic corrections that arise if the
$t\overline{b}H^-$ Yukawa coupling is expressed in terms of the $b$-quark pole
mass are automatically resummed, so that the convergence behaviour of the
perturbative expansion is improved.
On the other hand, the use of the $t$-quark $\overline{\rm MS}$ mass does not
entail such an improvement, but it rather appears somewhat unnatural from the
physical point of view.
The inclusion of the QCD corrections significantly reduces the theoretical
uncertainties due to scheme and typical scale variations.
We found that the QCD corrections to the total cross section may be of either
sign, depending on the values of $\sqrt s$, $\tan\beta$, and $m_H$, and that
they may reach a magnitude of up to 50\%.

The $e^+e^-$ LC TESLA, which is being developed at DESY, has a design
luminosity of $3.4\times10^{34}$~cm${}^{-2}$s${}^{-1}$
($5.8\times10^{34}$~cm${}^{-2}$s${}^{-1}$) at $\sqrt s=500$~GeV (800~GeV),
which corresponds to 340~fb${}^{-1}$ (580~fb${}^{-1}$) per year \cite{tdr}.
Thus, a total cross section of typically 0.03~fb (0.07~fb) will yield about
10 (40) signal events per year.

As a by-product of our analysis, we confirmed the numerical results for the
QCD-corrected total cross section of the SM process $e^+e^-\to q\overline{q}H$
obtained in Refs.~\cite{dit,daw}.
 
\bigskip

\noindent
{\bf Acknowledgements}

\smallskip

\noindent
We thank Stefan Dittmaier for providing us with numerical results from
Ref.~\cite{dit}, Thomas Hahn for helpful communications concerning
{\tt FormCalc} \cite{formcalc}, and Tao Han for a beneficial discussion
preceding our work.
F. M. thanks Stefan Berge for technical assistance regarding {\tt LoopTools}
\cite{formcalc}.
B.A.K. is grateful to the CERN Theoretical Physics Division for its
hospitality during a visit when this paper was finalized.
This work was supported in part by the Deutsche Forschungsgemeinschaft through
Grant No.\ KN~365/1-1, by the Bundesministerium f\"ur Bildung und Forschung
through Grant No.\ 05~HT1GUA/4, and by Sun Microsystems through Academic
Equipment Grant No.~EDUD-7832-000332-GER.

\renewcommand{\theequation}{\Alph{section}.\arabic{equation}}
\begin{appendix}
\setcounter{equation}{0} 

\section{\label{app:tree}Born form factors} 

In this Appendix, we list compact expressions for the form factors
$f_{\gamma\gamma}(x_1,x_2)$, $f_{\gamma Z}(x_1,x_2)$, and $f_{ZZ}(x_1,x_2)$
appearing in Eq.~(\ref{eq:born}).
It is possible to combine the propagators of the $t$ quark, $b$ quark, and
$H^-$ boson with their couplings to the photon and $Z$ boson by defining the
effective couplings
\begin{eqnarray}
{\cal Q}_q&=&-\frac{2c_ws_wQ_q}{1-x_q},\qquad
{\cal V}_q=\frac{I_q-2s_w^2Q_q}{1-x_q},\qquad
{\cal A}_q=\frac{I_q}{1-x_q},
\nonumber\\
{\cal H}_\gamma&=&\frac{2c_ws_w}{1-x_3},\qquad
{\cal H}_Z=\frac{s_w^2-c_w^2}{1-x_3},
\end{eqnarray}
where $q=t,b$ and we have identified $x_t=x_1$ and $x_b=x_2$.
The $t\overline{b}H^-$ coupling introduces $\tan\beta$ dependence through the
combinations
\begin{equation}
T_\pm=a_1\cot^2\beta\pm a_2\tan^2\beta.
\end{equation}

We find
\begin{eqnarray}
f_{Z Z}&=&
2{\cal V}_t^2\{4a_1a_2(2+2a_1-x_1)
+T_+[(1+2a_1)(1+a_1+a_2-a_3-x_1)-x_2(1-x_1)]\}
\nonumber\\&&{}
+4{\cal V}_t{\cal V}_b\{2a_1a_2(4+2a_1+2a_2-2a_3-x_1-x_2)
\nonumber\\&&{}
+T_+[(1+a_2-a_3-x_1)(1+a_1-a_3-x_2)+(a_1+a_2)(1+a_1+a_2-a_3)-a_1a_2]\}
\nonumber\\&&{}
+2{\cal V}_b^2\{4a_1a_2(2+2a_2-x_2)
+T_+[(1+2a_2)(1+a_1+a_2-a_3-x_2)-x_1(1-x_2)]\}
\nonumber\\&&{}
-4{\cal V}_t{\cal A}_tT_-[(1-x_2)(1+2a_1-x_1)-a_1+a_2-a_3]
\nonumber\\&&{}
+4{\cal V}_t{\cal A}_bT_-[(1+a_2-a_3-x_1)(1+a_1+2a_2-a_3-x_2)
\nonumber\\&&{}
+(a_1-a_2)(1+a_1+a_2-a_3)-3a_1a_2]
\nonumber\\&&{}
-4{\cal V}_b{\cal A}_tT_-[(1+a_1-a_3-x_2)(1+2a_1+a_2-a_3-x_1)
\nonumber\\&&{}
-(a_1-a_2)(1+a_1+a_2-a_3)-3a_1a_2]
\nonumber\\&&{}
+4{\cal V}_b{\cal A}_bT_-[(1-x_1)(1+2a_2-x_2)+a_1-a_2-a_3]
\nonumber\\&&{}
-2{\cal A}_t^2\{4a_1a_2(2+6a_1-3x_1)
-T_+[(1-6a_1)(1+a_1+a_2-a_3-x_1)-x_2(1-x_1)]\}
\nonumber\\&&{}
-4{\cal A}_t{\cal A}_b\{2a_1a_2(4+6a_1+6a_2-6a_3-3x_1-3x_2)
+T_+[(1+2a_1+a_2-a_3-x_1)
\nonumber\\&&{}
\times(1+a_1+2a_2-a_3-x_2)-(a_1+a_2)(1+a_1+a_2-a_3)+7a_1a_2]\}
\nonumber\\&&{}
-2{\cal A}_b^2\{4a_1a_2(2+6a_2-3x_2)
-T_+[(1-6a_2)(1+a_1+a_2-a_3-x_2)-x_1(1-x_2)]\}
\nonumber\\&&{}
+2{\cal V}_t{\cal H}_Z\{4a_1a_2(1-2a_1+2a_2-2a_3-x_2)
+T_+[(1+2a_2-2a_3-x_1)
\nonumber\\&&{}
\times(1-a_1-a_2-a_3-x_1-x_2)-2a_1(a_1+a_3)+2a_2(1+a_1+2a_2-2a_3)]\}
\nonumber\\&&{}
-2{\cal V}_b{\cal H}_Z\{4a_1a_2(1+2a_1-2a_2-2a_3-x_1)
+T_+[(1+2a_1-2a_3-x_2)
\nonumber\\&&{}
\times(1-a_1-a_2-a_3-x_1-x_2)-2a_2(a_2+a_3)+2a_1(1+2a_1+a_2-2a_3)]\}
\nonumber\\&&{}
-2{\cal A}_t{\cal H}_ZT_-[(1+2a_1+2a_2-2a_3-x_1)(1+a_1-a_2-a_3-x_1-x_2)
\nonumber\\&&{}
+2a_2(1-2a_1+2a_2-2a_3)]
\nonumber\\&&{}
-2{\cal A}_b{\cal H}_ZT_-[(1+2a_1+2a_2-2a_3-x_2)(1-a_1+a_2-a_3-x_1-x_2)
\nonumber\\&&{}
+2a_1(1+2a_1-2a_2-2a_3)]
\nonumber\\&&{}
-{\cal H}_Z^2[4a_1a_2(3-4a_3-2x_1-2x_2) 
\nonumber\\&&{}
+T_+(3-4a_3-2x_1-2x_2)(1+a_1+a_2-a_3-x_1-x_2)].
\label{eq:zz}
\end{eqnarray}
The formulas for $f_{\gamma\gamma}(x_1,x_2)$ and $f_{\gamma Z}(x_1,x_2)$ may
be obtained from Eq.~(\ref{eq:zz}) by adjusting the coupling constants.
Specifically, one has to substitute
\begin{equation}
{\cal V}_q\to{\cal Q}_q,\qquad
{\cal A}_q\to0,\qquad
{\cal H}_Z\to{\cal H}_\gamma
\end{equation}
in the first case and
\begin{eqnarray}
{\cal V}_q{\cal V}_{q^\prime}&\to&
{\cal Q}_q{\cal V}_{q^\prime}+{\cal Q}_{q^\prime}{\cal V}_q,\qquad
{\cal V}_q{\cal A}_{q^\prime}\to{\cal Q}_q{\cal A}_{q^\prime},\qquad
{\cal A}_q{\cal A}_{q^\prime}\to0,
\nonumber\\
{\cal V}_q{\cal H}_Z&\to&
{\cal Q}_q{\cal H}_Z+{\cal V}_q{\cal H}_\gamma,\qquad
{\cal A}_q{\cal H}_Z\to{\cal A}_q{\cal H}_\gamma,\qquad
{\cal H}_Z^2\to2{\cal H}_\gamma{\cal H}_Z
\end{eqnarray}
in the second one ($q,q^\prime=t,b$).

\section{\label{app:four}Four-particle phase space}

In this Appendix, we present the parameterization of the four-particle phase
space that we use to evaluate the hard-gluon contribution.
We generically denote the four-momenta and masses of the final-state particles
as $p_i$ and $m_i$ ($i=1,\ldots,4$), respectively.
Similarly as in Sec.~\ref{sec:ana}.1, we define $p=\sum_{i=1}^4p_i$, $s=p^2$,
$a_i=m_i^2/s$, $x_i=2p\cdot p_i/s$, and $y_i=\sqrt{x_i^2-4a_i}$.
Due to four-momentum conservation, we have $\sum_{i=1}^4x_i=2$.
We decompose the four-particle phase space into three nested two-particle
phase spaces as \cite{byk}
\begin{equation}
d{\rm PS}_4(p;p_1,p_2,p_3,p_4)=\frac{ds_{12}ds_{34}}{(2\pi)^2}
d{\rm PS}_2(p;p_{12},p_{34})\,
d{\rm PS}_2(p_{12};p_1,p_2)\,d{\rm PS}_2(p_{34};p_3,p_4),
\end{equation}
where $p_{ij}=p_i+p_j$ and $s_{ij}=p_{ij}^2$, with $(i,j)=(1,2),(3,4)$.
As in Sec.~\ref{sec:ana}.1, we work in the c.m.\ frame, take the $z$ axis of
the coordinate system to point along
$\mbox{\boldmath$k_1$}=-\mbox{\boldmath$k_2$}$, and choose the $x$ axis
arbitrarily.
We have $\mbox{\boldmath$p_{12}$}=-\mbox{\boldmath$p_{34}$}$ and
$|\mbox{\boldmath$p_{12}$}|=(1/2)\sqrt{\lambda(s,s_{12},s_{34})/s}$.
Using
\begin{eqnarray}
d{\rm PS}_2(p;p_{12},p_{34})
&=&\frac{|\mbox{\boldmath$p_{12}$}|}{(4\pi)^2\sqrt{s}}d\cos\theta_{12}\,
d\phi_{12},
\nonumber\\
d{\rm PS}_2(p_{ij};p_i,p_j)
&=&\frac{1}{(4\pi)^2|\mbox{\boldmath$p_{ij}$}|}dp_i^0\,d\phi_i,
\end{eqnarray}
where $\theta_{12}$ and $\phi_{12}$ are the polar and azimuthal angles of
$\mbox{\boldmath$p_{12}$}$, respectively, and $\phi_i$ is the azimuthal angle
of $\mbox{\boldmath$p_i$}$ with respect to the axis pointing along
$\mbox{\boldmath$p_{ij}$}$ measured from the plane spanned by
$\mbox{\boldmath$k_1$}$ and $\mbox{\boldmath$p_{ij}$}$, we obtain
\begin{equation}
d{\rm PS}_4(p;p_1,p_2,p_3,p_4)
=\frac{8}{(4\pi)^8\sqrt{\lambda(s,s_{12},s_{34})}}ds_{12}\,ds_{34}\,
d\cos\theta_{12}\,d\phi_{12}\,dp_1^0\,d\phi_1\,dp_3^0\,d\phi_3.
\end{equation}
Owing to the azimuthal symmetry of the problem under consideration, the
integration over $\phi_{12}$ is trivial, and we may choose
$\mbox{\boldmath$p_{12}$}$ to lie in the $x$-$z$ plane.
If we now rotate the coordinate system in such a way that
$\mbox{\boldmath$p_{12}$}$ points along the $z$ axis and
$\mbox{\boldmath$p_1$}$ lies in the $x$-$z$ plane, then $\theta=\theta_{12}$
and $\phi=\pi-\phi_1$ define the direction of $\mbox{\boldmath$k_1$}$.
Introducing $z_{ij}=s_{ij}/s$, we thus have 
\begin{equation}
d{\rm PS}_4(p;p_1,p_2,p_3,p_4)
=\frac{s^2}{(4\pi)^7\sqrt{\lambda(1,z_{12},z_{34})}}
dz_{12}\,dz_{34}\,dx_1\,dx_3\,d\phi_3\,d\cos\theta\,d\phi.
\end{equation}
As explained in Eq.~(\ref{eq:int}), the integrations over $\cos\theta$ and
$\phi$ can be exploited to transform the leptonic tensor $L^{\mu\nu}$ defined
by Eq.~(\ref{eq:lep}), which depends on $k_1$ and $k_2$, into one depending
only on $p$, so as to preclude scalar products of the type $k_i\cdot p_j$.
The residual scalar products read
\begin{eqnarray}
p_1\cdot p_2&=&\frac{s}{2}(z_{12}-a_1-a_2),
\nonumber\\
p_3\cdot p_4&=&\frac{s}{2}(z_{34}-a_3-a_4),
\nonumber\\
p_1\cdot p_3&=&\frac{s}{4}[x_1x_3-y_1y_3(\sin\theta_1\sin\theta_3\cos\phi_3
-\cos\theta_1\cos\theta_3)],
\nonumber\\
p_1\cdot p_4&=&\frac{s}{2}(x_1-z_{12}-a_1+a_2)-p_1\cdot p_3,
\nonumber\\
p_2\cdot p_3&=&\frac{s}{2}(x_3-z_{34}-a_3+a_4)-p_1\cdot p_3,
\nonumber\\
p_2\cdot p_4&=&\frac{s}{2}(1-x_1-x_3+a_1-a_2+a_3-a_4)+p_1\cdot p_3,
\end{eqnarray}
where $\theta_i$ is the angle enclosed between $\mbox{\boldmath$p_{ij}$}$ and
$\mbox{\boldmath$p_i$}$.
It is determined by
\begin{equation}
\cos\theta_i=\frac{x_i(1+z_{ij}-z_{kl})-2(z_{ij}+a_i-a_j)}
{y_i\sqrt{\lambda(1,z_{ij},z_{kl})}},
\end{equation}
with $(i,j),(k,l)=(1,2),(3,4)$ and $(i,j)\ne(k,l)$.
Furthermore, we have
\begin{equation}
x_j=1-x_i+z_{ij}-z_{kl}.
\end{equation}

The limits of integration are
\begin{eqnarray}
(\sqrt{a_1}+\sqrt{a_2})^2&<&z_{12}<\left(1-\sqrt{a_3}-\sqrt{a_4}\right)^2,
\nonumber\\
(\sqrt{a_3}+\sqrt{a_4})^2&<&z_{34}<(1-\sqrt{z_{12}})^2,
\nonumber\\
x_i^-&<&x_i<x_i^+,
\nonumber\\
0&<&\phi_3<2\pi,
\label{eq:lim}
\end{eqnarray}
where
\begin{equation}
x_i^\pm=\frac{1}{2z_{ij}}\left[(1+z_{ij}-z_{kl})(z_{ij}+a_i-a_j)
\pm\sqrt{\lambda(1,z_{ij},z_{kl})\lambda(z_{ij},a_i,a_j)}\right].
\end{equation}

For the application in Sec.~\ref{sec:ana}.3, it is convenient to assign the
indices 1, 2, 3, and 4 to the $t$ quark, $\overline{b}$ quark, gluon, and
$H^-$ boson, respectively.
Then, the hard-gluon condition $q^0>E_{\rm cut}$ may be implemented by
substituting $x_3^-\to\max\left(x_3^-,2E_{\rm cut}/\sqrt s\right)$ in
Eq.~(\ref{eq:lim}).
Furthermore, we have $a_3=0$ throughout this appendix.

\end{appendix}


\end{document}